\documentclass[preprint]{aastex}
\usepackage{apjfonts}
\usepackage{graphicx}
\usepackage{psfig}
\usepackage{float}

\usepackage{graphics}
\usepackage{longtable}
\usepackage{amssymb,amsmath}
\usepackage{lscape}

\graphicspath{{Figures/}}

\newcommand{\degree}{\ensuremath{^\circ}}

\newcommand{\msun}{\ensuremath{\rm{M}_\odot}}

\newcommand{\kms}{\ensuremath{\rm{km \: s}^{-1}}}

\newcommand{\sneia}{SNe Ia}
\newcommand{\snia}{SN Ia}

\newcommand{\WThreeeSigma}{\ensuremath{ 0.067} \AA}
\newcommand{\WMattila}{\ensuremath{ 5.8} \AA}
\newcommand{\WPlotted}{\ensuremath{ 0.12} \AA}

\newcommand{\MassLimitPlotted}{\ensuremath{ 0.001} \msun}
\newcommand{\MassLimit}{\ensuremath{ 5.8 \times 10^{-4}} \msun}

\newcommand{\Binning}{\ensuremath{ 4.0} \AA}
\newcommand{\SmoothWidth}{\ensuremath{           60} \AA}
\newcommand{\OneSigma}{\ensuremath{ 0.0024} \AA}
\newcommand{\PeakFluxMattila}{\ensuremath{ 6.71 \times 10^{-17}} erg s\ensuremath{^{-1}} cm\ensuremath{^{-2}} \AA\ensuremath{^{-1}}}

\newcommand{\FluxLimitPlotted}{\ensuremath{ 3.14 \times 10^{-17}} erg s\ensuremath{^{-1}} cm\ensuremath{^{-2}}}
\newcommand{\FluxLimitModel}{\ensuremath{ 2.20 \times 10^{-17}} erg s\ensuremath{^{-1}} cm\ensuremath{^{-2}}}

\newcommand{\FluxLimitSNPlotted}{\ensuremath{ 1.57 \times 10^{35}} erg s\ensuremath{^{-1}}}

\newcommand{\Angle}{\ensuremath{ 30.0}}

\begin{document}

\title{No Stripped Hydrogen in the Nebular Spectra of Nearby Type Ia
Supernova 2011fe\footnote{Based on data acquired using the Large
Binocular Telescope (LBT/MODS).}}  
\shorttitle{No Stripped Hydrogen in SN 2011fe} 
\shortauthors{Shappee, Stanek, Pogge \& Garnavich}

\author{
{Benjamin J. ~Shappee}\altaffilmark{2,3}, 
{K. Z. ~Stanek}\altaffilmark{2,4},
{R. W. ~Pogge}\altaffilmark{2,4},
and 
{P. M. ~Garnavich}\altaffilmark{5},
}

\email{shappee@astronomy.ohio-state.edu, kstanek@astronomy.ohio-state.edu, pogge@astronomy.ohio-state.edu, pgarnavi@nd.edu}

\altaffiltext{2}{Department of Astronomy, The Ohio State University, Columbus, Ohio 43210, USA}
\altaffiltext{3}{NSF Graduate Fellow}
\altaffiltext{4}{Center for Cosmology \& AstroParticle Physics, The Ohio State University, Columbus, OH 43210, USA}
\altaffiltext{5}{Department of Physics, University of Notre Dame, Notre Dame, IN 46556, USA}

\date{\today}

\begin{abstract}

A generic prediction of the single-degenerate model for Type
Ia supernovae (\sneia{}) is that a significant amount of material
will be stripped from the donor star ($\sim$0.5 \msun{} for a giant donor
and $\sim$0.15 \msun{} for a main sequence donor) by the supernova ejecta. This
material, excited by gamma-rays from radioactive decay, would then
produce relatively narrow ($\lesssim$1000 \kms) emission features
observable once the supernova enters the nebular phase. Such emission
has never been detected, which already provides strong constraints on
 Type Ia progenitor models.  In this Letter we report the deepest
limit yet on the presence of H$\alpha$ emission originating from the
stripped hydrogen in the nebular spectrum of a Type Ia supernova
obtained using a high signal-to-noise spectrum of the nearby
normal \snia{} 2011fe 274 days after $B$-band maximum light with the Large Binocular Telescope's Multi-Object Double Spectrograph. We put
a conservative upper limit on the H$\alpha$ flux of \FluxLimitPlotted{}, which corresponds to a luminosity of
\FluxLimitSNPlotted. By scaling models from the literature, our flux limit translates into an upper limit of
$\lesssim$\MassLimitPlotted{} of stripped material. This is an order of magnitude stronger than previous limits. SN 2011fe was a 
typical Type Ia supernova, special only in its proximity, and we
argue that lack of hydrogen emission in its nebular spectrum adds yet
another strong constraint on the single degenerate class of models for
\sneia{}.

\end{abstract}

\section{Introduction}
\label{sec:introduc}

Despite the fact that \sneia{} were used to discover the accelerating
universe \citep{riess98, perlmutter99}, the physical nature of their
progenitor systems remains theoretically ambiguous and observationally
elusive (for a review see \citealp{wang12}). It is commonly accepted
that \sneia{} result from the thermonuclear explosion of a
carbon-oxygen white dwarf (WD) in a close binary system, but the
nature of the binary companion and the sequence of events leading to
the SN explosion are still uncertain.  There are two dominant models: the double degenerate (DD) scenario, in which the
companion is also a WD \citep{tutukov79, iben84, webbink84}, and the single
degenerate (SD) scenario, in which the companion is a non-degenerate
object such as a main sequence (MS) star, a red giant (RG), a sub-giant or a
He star \citep{whelan73, nomoto82}.  There are additional minor variations on
these basic models of the progenitor system for both the SD channel
(e.g.~\citealp{justham11, wheeler12}) and the DD channel
(e.g.~\citealp{thompson11, shappee12a}). These uncertainties about the 
progenitor systems for \sneia{} remain a substantial problem for understanding the 
systematic errors in using \sneia{} to study cosmology (e.g.~\citealp{wood-vasey07}).

One of the observational signatures of the SD model is that the material
from the companion should be stripped when struck by the SN ejecta
\citep{wheeler75}, leading to both immediate signatures from the impact of the SN ejecta on the companion (e.g.~\citealp{marietta00, meng07, pakmor08,
pan12, liu12}) and major changes in the companion's future evolution
\citep{podsiadlowski03, shappee12b, pan12b}.  Recently, the hydrodynamic simulations of \citet{pan12} and \citet{liu12} have shown that $\sim 0.1 \-- 0.2$ \msun{} of solar-metallicity material is expected to be removed from MS companions by the impact of the SN ejecta.  
These hydrodynamic simulations show that this material will be embedded in low-velocity supernova
debris with a characteristic velocity of $\lesssim$1000 \kms{}.  However, the line profiles from the stripped material are somewhat uncertain because this material will be asymmetrically stripped \citep{liu12}, and the orientation of the binary relative to our line of sight, at the time of explosion, is unknown. This material will be hidden in early-time spectra by higher
velocity, optically thick, iron-rich ejecta, but will then appear in
late-time, nebular phase spectra ($\gtrsim 250$ days;
\citealp{mattila05}) as the higher velocity ejecta become optically
thin.  
                                    
Only a handful of nebular phase, high signal-to-noise (S/N) \sneia{} spectra have been
published in the literature, with the strongest limits on late-time
hydrogen flux coming from \citet{mattila05} and \citet{leonard07}.
\citet{mattila05} obtained late-time low-resolution spectra of SN
2001el, modeled the emission from solar-metallicity material stripped
from a non-degenerate companion, and used both to place an upper limit
of $\lesssim$0.03 \msun{} on the presence of this material.
\citet{leonard07} obtained deep, medium-resolution, nebular-phase,
multi-epoch spectra of SN 2005am and SN 2005cf at distances $\sim$37
Mpc and $\sim$32 Mpc, respectively.  Using the approach of
\citet{mattila05}, he placed a $\lesssim$0.01 \msun{} upper limit on the presence of any low velocity solar-abundance material in both SN. These
observations seem to firmly rule out main sequence companions for
these three SNe.

However, a more exotic scenario which might evade these constraints has since been proposed by \citet{justham11}.  In this model a WD is spun up by the mass it accretes from its
non-degenerate companion, allowing the WD to remain stable above the
Chandrasekhar mass and giving the companion time to evolve and contract before the SN explosion.  This leaves a smaller and more tightly bound 
companion at the time of explosion, reducing the
amount of stripped material. It therefore is of significant
interest to put even stronger constraints on the amount of stripped material for other
\sneia{}.

At a mere 6.4 Mpc \citep{shappee11} the ``plain vanilla'' Type Ia SN (as described by \citealp{wheeler12}) is an ideal target for obtaining improved constraints. SN 2011fe was discovered less than one day after explosion by
the Palomar Transient Factory \citep{law09} in the Pinwheel Galaxy
(M101), a well-studied, nearby face-on spiral.  This SN is the nearest Ia in
the last 25 years and the early detection and announcement allowed extensive
multi-wavelength follow-up observations \citep{nugent11}.  Additionally, SN 2011fe is only slightly reddened
and is surrounded by a relatively ``clean'' environment \citep{patat11}.
\citet{brown12b} and \citet{bloom12} used the very early-time UV
and optical observations to constrain the shock heating that would
occur as the SN ejecta collide with the companion, ruling out giants
and main sequence secondaries more massive than the Sun.  These
constraints appear to rule out SD models; however, these studies both
rely on \citet{kasen10}, which assumes a Roche-lobe overflowing
secondary with a typical stellar structure and depends on the orientation of the binary relative to our line of sight. However, \citet{meng07} pointed out that the secondary will be left smaller and more
compact than a typical main sequence star after binary evolution, which would produce a more subtle photometric signal at
even earlier times which could have been missed by early-time
observations. 

In this study we place the strongest limits yet on the presence of H$\alpha$ emission in the nebular spectrum of a Type Ia
supernova (\snia{}).  We obtained a very high S/N spectrum of \snia{} 2011fe 274 days after maximum $B$-band light using the Multi-Object Double Spectrograph on the 8.4-meter Large Binocular Telescope.  We generally followed the analysis procedures used by \citet{leonard07}, so our constraints should be directly comparable to earlier results. In
Section \ref{sec:Obs} we describe our observations, in Section
\ref{sec:Hydrogen} we derive our H$\alpha$ limits, and we summarize our
findings in Section \ref{sec:conclusion}.

\section{Observations}
\label{sec:Obs}

We obtained five high-S/N spectra of SN 2011fe from 73 to 274
days after maximum $B$-band light using the first of the Multi-Object
Double Spectrographs (MODS; \citealp{pogge10}) on the Large Binocular
Telescope (LBT).  The two channels of MODS allowed us to obtain wide
spectral coverage ($3200 \-- 10000 $\AA) in a single exposure, which will make these observations useful for many future studies.  Details of the observations
and the flux standards used are listed in Table \ref{tab:1}.  All observations were taken through a $1\arcsec$ wide slit.  To perform
basic CCD reductions on our spectra we followed the ``MODS Basic CCD
Reduction with modsCCDRed'' manual\footnote{modsCCDRed manual is
available here
\url{http://www.astronomy.ohio-state.edu/MODS/Manuals/MODSCCDRed.pdf}.}.  We
then performed cosmic ray rejection using L.A.Cosmic
\citep{vandokkum01} and combined each channel's 2D spectra for each
epoch. Next, we extracted the 1D sky-subtracted spectra using the {\it
apall} task in IRAF.\footnote{IRAF is distributed by the National
Optical Astronomy Observatory, which is operated by the Association of
Universities for Research in Astronomy (AURA) under cooperative
agreement with the National Science Foundation.}  Each spectrum was
then wavelength and flux calibrated.  We did not attempt to correct
small-scale telluric absorption bands as performed by
\citet{leonard07}.  This will not adversely affect our results because
the telluric absorptions occurs on wavelength scales smaller than the
H$\alpha$ emission expected from the SN.  This also avoids adding any noise from telluric spectra to our SN spectra.

\begin{deluxetable}{lrrrcrccccc}
\tabletypesize{\tiny}
\tablewidth{450pt}
\tablecaption{Observations}
\tablehead{\colhead{} & 
\colhead{} &
\colhead{HJD} &
\colhead{P.A.} &
\colhead{Par. P.A.} &
\colhead{} & 
\colhead{Flux } &
\colhead{Seeing} &
\colhead{Exposure} &
\colhead{$r^{\prime}$} &
\colhead{Scale} \\
\colhead{UT Date} &
\colhead{Day} &
\colhead{$-$2,400,000} &
\colhead{(deg)} &
\colhead{(deg)} &
\colhead{Airmass} &
\colhead{standards} &
\colhead{(arcsec)} &
\colhead{(s)}  &
\colhead{(mag)}  &
\colhead{factor}  }
\startdata

2011 Nov 23.52 & 73.02  & 55889.02  &  $-78.0$ & $-84.3$ -- $-85.7$ & 1.60 -- 1.64 & Feige110, G191-B2B  & 0.8 / 0.9 &  360.0 / 360.0 & \nodata\tablenotemark{a} & 1.07\\

2012 Jan 2.54 & 113.05 & 55929.05 & $-124.4$ & $-134$ -- $-139$ & 1.11 -- 1.12 & Feige34 & 0.8 / 0.9 & 900.0 / 900.0 & $14.26 \pm 0.03$& 1.93 \\

2012 Mar 24.47 & 194.97 & 56010.97  & $-68.5$ & $110$ -- $124$ & 1.15 -- 1.24 & Feige67 & 0.5 / 0.7 & 2880.0 / 2160.0 & $16.35 \pm 0.01$ & 1.77 \\

2012 Apr 27.27 & 228.77 & 56044.77 & $-2.0$ & $-159$ -- $177$ & 1.07 -- 1.08 & HZ44, BD+33d2642  & 0.9 / 1.1 & 3600.0 / 3600.1 & $17.06 \pm 0.01$ & 2.20 \\

2012 Jun 12.16 & 274.66 & 56090.66 & $-32.7$ & $-177$ -- $129$ & 1.07 -- 1.13 & HZ44 & 0.7 / 0.7 & 7200.0 / 7200.0 & $17.95 \pm 0.01$ & 1.26 \\

\enddata \tablecomments{Observational and derived properties of our time series spectra of SN 2011fe. Days since maximum $B$ brightness assume JD $t_{B {\rm max}} = 2455816.0 \pm 0.3$ \citep{richmond12}. P.A. is the position angle of the spectrograph slit.  Par. P.A. and airmass give the range of parallactic angles and airmasses at the start of each separate observation, respectively. Standard stars were observed on the same night as science observations and, when multiple standards were available, the computed response functions from each standard were averaged. Seeing gives the FWHM for the red / blue channels of the spatial profile in the combined 2d spectra. Exposure times are given for the red / blue channels separately. $r^{\prime}$ magnitude is derived from the acquisition images. Each flux calibrated spectra was multiplied by a scale factor derived in the $R$- or $r^{\prime}$-band to place it on an absolute flux scale.}

\tablenotetext{a}{SN 2011fe was saturated in this epoch's acquisition images, so no reliable photometry was obtained.}

\label{tab:1}

\end{deluxetable}

\begin{figure}[htp]
	\includegraphics[width=16cm]{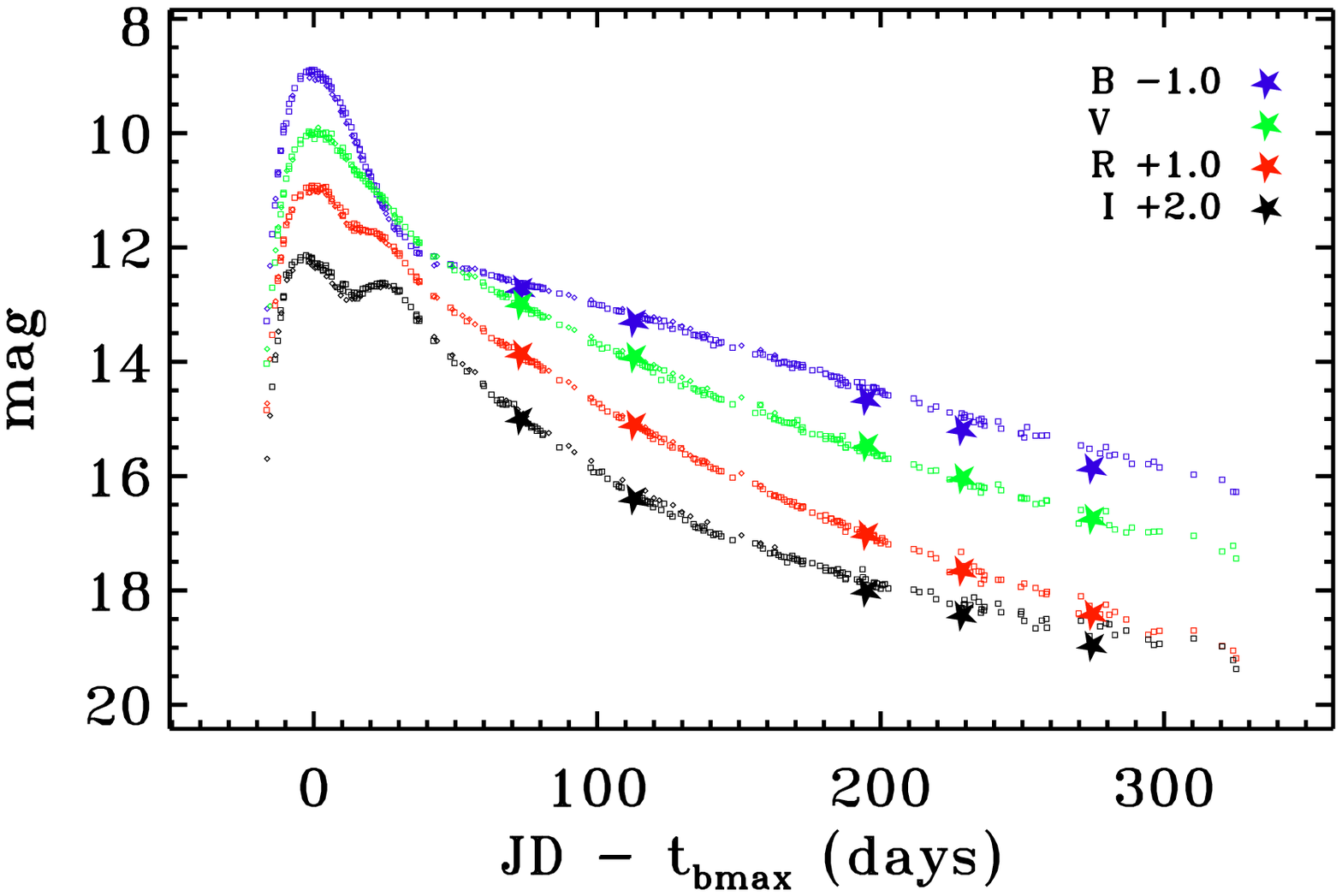}

	\caption{SN 2011fe light curves relative to the time of maximum
	$B$-band brightness ($t_{B {\rm max}} = 2455816.0 \pm 0.3$, 
	\citealp{richmond12}).  Open diamonds and open squares show the $BVRI$ photometry from \citet{richmond12} and \citet{munari12}, respectively.  Our synthetic
	$BVRI$ photometry are represented by filled stars. The first
	epoch is calibrated to an absolute flux scale using the
	\citet{richmond12} $R$-band light curve so good agreement is
	expected.  The rest of our spectra are calibrated using our
	$r'$-band acquisition images and SDSS DR7 photometry.  From
	the second epoch we estimate that our absolute flux
	calibration is accurate to better than $10 \%$ in the
	$R$-band.}
	\label{fig:lightcurve}
\end{figure}
\begin{figure}[htp]
	\includegraphics[width=16cm]{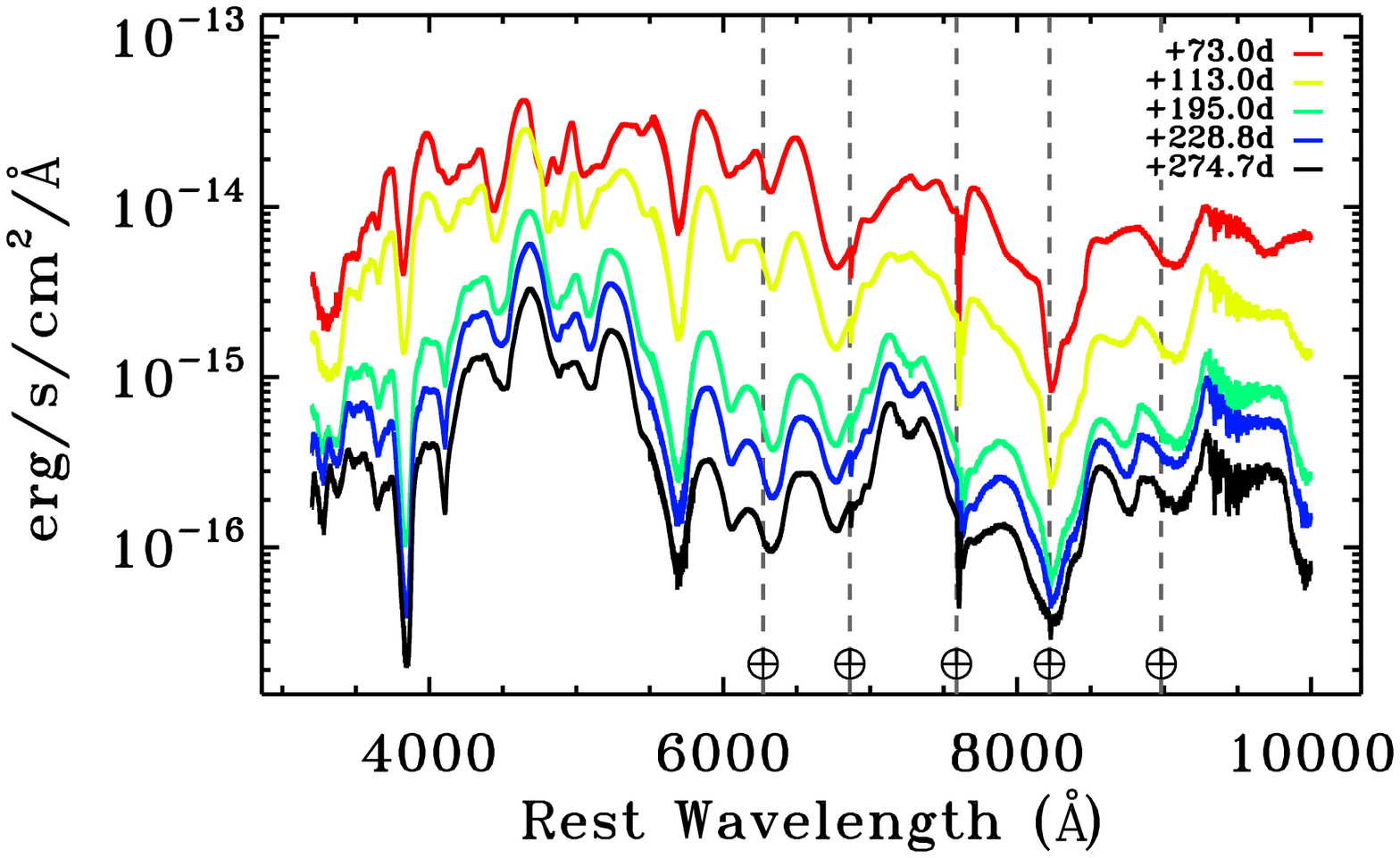}

	\caption{Rest-frame absolute-flux-calibrated spectra of SN 2011fe.  Vertical dashed gray lines mark obvious telluric
	features.  Spectra are labeled by days since maximum $B$
	brightness.}
	\label{fig:spectra}
\end{figure}

Because our spectra were taken under non-photometric conditions with a
relatively narrow ($1\arcsec$) slit, it is necessary to scale the fluxes of our spectra to
place them on an absolute flux scale.  To do this we performed
aperture photometry on our Sloan $r'$ \citep{fukugita96} acquisition
images using the IRAF package {\it apphot}.  Because the SN was
saturated in the acquisition image of the first epoch we must treat
that spectrum separately.  For the remaining epochs we used the four
brightest stars in the image to put our photometry on the Sloan
Digital Sky Survey (SDSS; \citealp{york00}) Data Release 7
\citep{abazajian09} magnitude system. Our $r'$-band magnitudes are
reported in Table \ref{tab:1}.  We then scale the spectrum so that its
synthetic $r'$-band photometry matches the computed $r'$-band aperture 
photometry. For the first epoch we scaled the spectrum such that a synthetic $R$-band magnitude 
\citep{bessell12} matches the $R$-band magnitude for its epoch found by interpolating the combined $R$-band light curves of \citet{richmond12} and \citet{munari12}.  To convert from the \textit{AB} magnitude
system \citep{oke74} of our synthetic photometry to the Vega magnitude
system reported in \citet{richmond12} and \citet{munari12}, we use the conversion presented
in \citet{blanton07}.  The factor by which each spectrum was
multiplied is reported in Table \ref{tab:1}.  From the last four epochs'
spectra, which are calibrated from the acquisition image but are also
covered by the combined $R$-band light curves of \citet{richmond12} and \citet{munari12}, we estimate
that our absolute flux calibration in the $R$-band (where H$\alpha$ is
located) is accurate to $10 \%$ or better.  In Figure
\ref{fig:lightcurve} we show the \textit{BVRI} light
curves of \citet{richmond12} and \citet{munari12}  as compared to our synthetic \textit{BVRI} photometry from the spectra for illustrative purposes.  Figure \ref{fig:spectra} presents
our calibrated spectra with obvious telluric features marked.
Finally, we compare our nebular phase spectrum to the spectra
presented in \citet{leonard07}, which provided the previous best limits
on $H\alpha$.  As can be seen in Figure \ref{fig:spectra}, our
spectrum has a substantially higher S/N than the previous studies.  These observations allow us to place even stronger
limit on H$\alpha$ emission as we discuss in \S\ref{sec:Hydrogen}.

\begin{figure}[htp]
	\includegraphics[width=16cm]{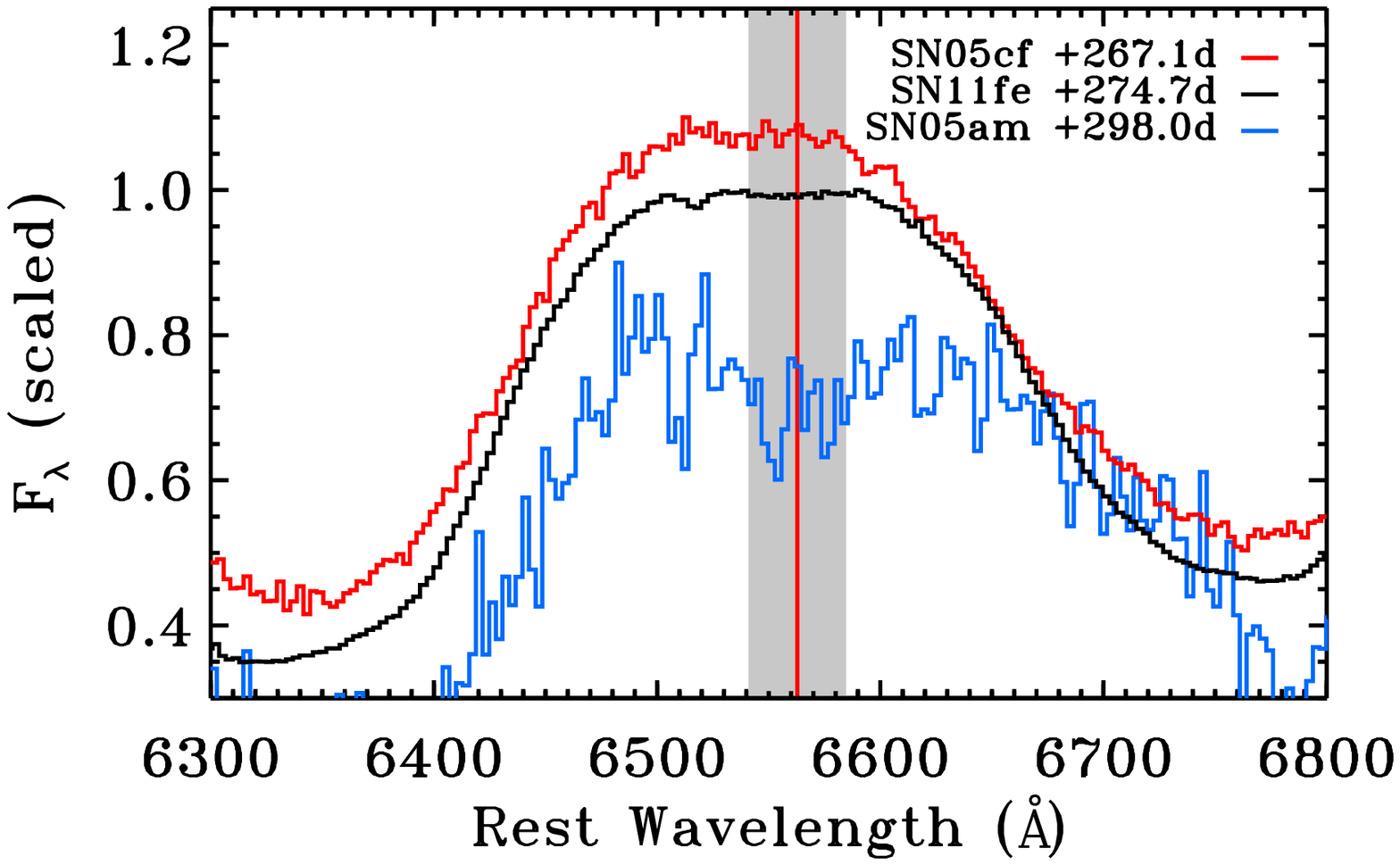}

	\caption{Comparison between our nebular phase spectrum of SN
	2011fe and the spectra of SN 2005cf and SN 2005am presented
	in \citet{leonard07} which previously provided the best limits
	on $H\alpha$ emission from the stripped material. Our spectra
	are binned to match the resolution of \citet{leonard07}.  The rest wavelength of H$\alpha$ is indicated by the vertical red line and the
	shaded gray region shows where hydrogen emission would be
	expected ( $\pm 1000$ \kms{} $= \pm 22 $ \AA{} about H$\alpha$). Our spectrum has a substantially
	higher S/N due to the proximity of SN 2011fe, allowing us to place a substantially stronger
	limit on H$\alpha$ emission in Section \ref{sec:Hydrogen}. }
	\label{fig:leonard_comp}
\end{figure}

\section{The Search for Hydrogen}
\label{sec:Hydrogen}

In our search for H$\alpha$ emission in the spectrum of SN 2011fe, we
closely follow the methods presented in \citet{leonard07}.  We define a continuum by smoothing our spectra on scales large compared to expected H$\alpha$ feature, then subtract off this continuum and examine the residuals. 
We searched for H$\alpha$ emission within $\pm 1000$ \kms{} ($\pm 22 $
\AA) about H$\alpha$ at the redshift of M101, $0.000804 \pm 0.000007$
\citep{devaucouleurs91}.  We smoothed the spectrum with a
second-order Savitsky-Golay smoothing polynomial \citep{press92} with
a width of \SmoothWidth{}. Our data requires this smaller smoothing
scale than that employed by \citet{leonard07} for a good continuum
fit.  It is, however, still significantly larger than the expected
velocity width of any H$\alpha$ feature.  Additionally, we smoothed over the telluric feature at $6510 \-- 6525$~\AA{} before
smoothing the rest of the spectrum, because the feature was affecting the
continuum determination near H$\alpha$.

\begin{figure}[htp]
	\includegraphics[width=12cm]{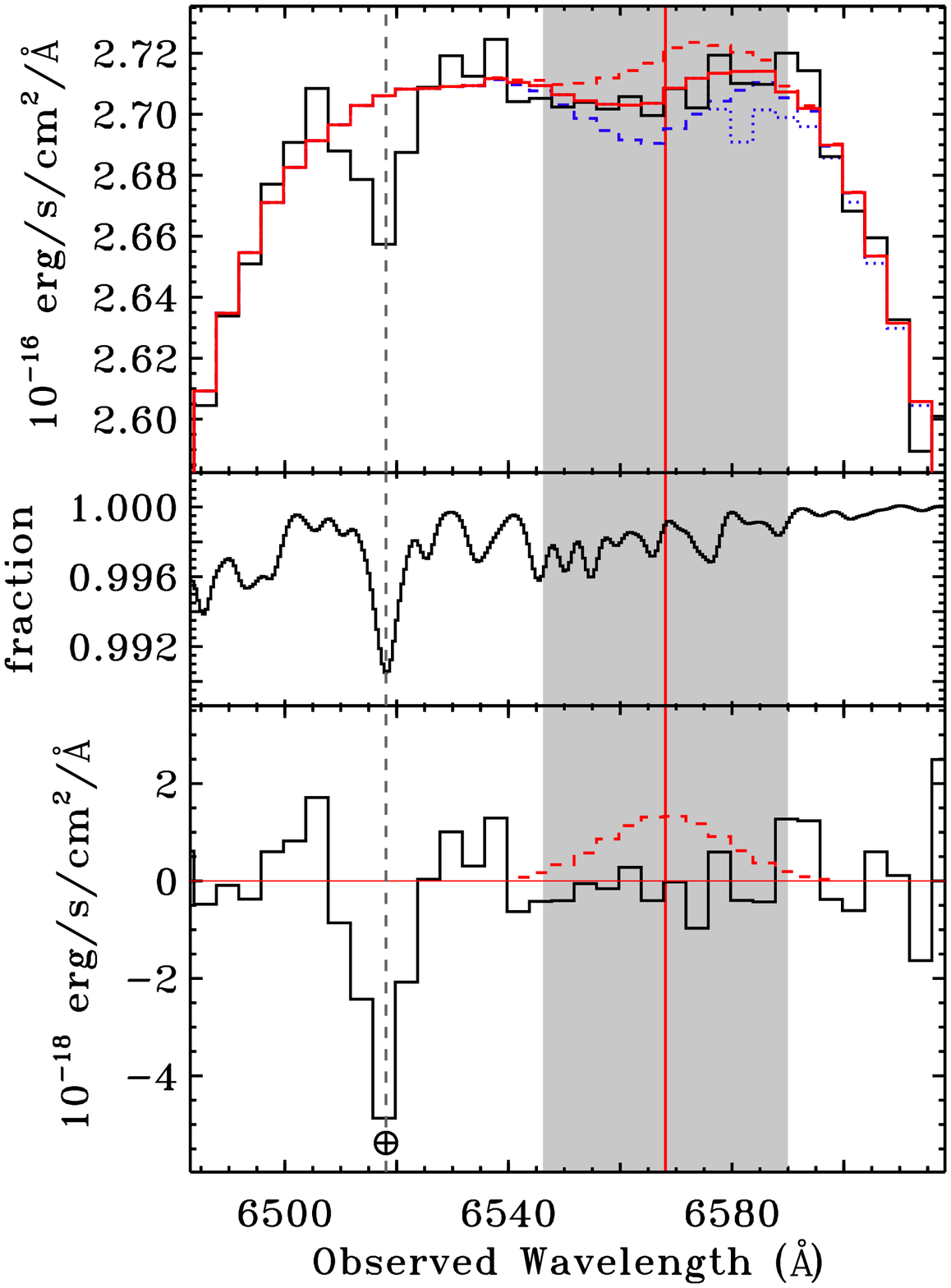}

	\caption{Nebular phase spectrum of SN 2011fe illustrating our conservative
	limit of \FluxLimitPlotted{} on the flux of H$\alpha$ emission. The rest wavelength of H$\alpha$ is indicated by the vertical red line and the
	shaded gray region shows where hydrogen emission would be
	expected ( $\pm 1000$ \kms{} $= \pm 22 $ \AA{} about H$\alpha$).  Adopting the models of \citet{mattila05}, these
	limits translate into a $\lesssim$\MassLimitPlotted{} limit on the
	amount of solar-abundance material stripped from the
	companion.  Vertical dashed gray line marks a large telluric
	feature.  {\it Top Panel:} SN spectrum binned to the
	approximate spectral resolution (\Binning{}; black solid);
	smoothed continuum (solid red); smoothed continuum with
	H$\alpha$ limit added (dashed red); smoothed continuum
	with H$\alpha$ limit subtracted to show what the underlying
	continuum would have to be to get the observed spectrum
	(dashed blue); and smoothed continuum
	with H$\alpha$ limit subtracted, assuming the velocity distribution discussed in Section \ref{sec:ConLimit} (dotted blue).  {\it Middle Panel:} The location of telluric water
	vapor absorption lines illustrated by the ESO SM-01 Sky Model Mode Version 1.3.1. {\it Bottom Panel:} SN spectrum with
	smoothed continuum subtracted (solid black) as compared to the  H$\alpha$ limit
	(dashed red).   Horizontal solid
	red line marks zero.}
	\label{fig:halpha}
\end{figure}

  Our nebular phase spectrum
and continuum fit are shown in the top panel of Figure
\ref{fig:halpha} in the vicinity of H$\alpha$ and are binned to their
approximate spectral resolution.

We then subtracted the continuum from the binned spectrum and examined
the residuals, shown in the bottom panel of Figure \ref{fig:halpha},
for narrow H$\alpha$ emission.  There is no evidence for any H$\alpha$
emission in the spectrum.  The closest possible emission is the hump in the smoothed continuum at $\sim$6575$\--$6595 \AA{} which is discussed in Section \ref{sec:ConLimit}. More broadly, we found no narrow emission lines at any wavelength,
including regions around H$\beta$, [\ion{O}{1}] $\lambda\lambda 6300,
6364$, [\ion{O}{2}] $\lambda\lambda 7319, 7330$, [\ion{O}{3}]
$\lambda\lambda 4959, 5007$, [\ion{Ca}{2}] $\lambda\lambda 7291, 7324$
, and [\ion{N}{2}] $\lambda\lambda 6548, 6583$.

\subsection{Statistical Limit}
\label{sec:StatLimit}

Following \citet{leonard01} and \citet{leonard07}, we compute a 
$3\sigma$ upper bound on the equivalent width as
\begin{eqnarray}
  W_\lambda(3\sigma) &=& 3 \Delta\lambda\ \Delta I \sqrt{W_{\textrm{line}} / \Delta\lambda\ } \sqrt{1 / B}
  \nonumber \\
  &=& 3 \Delta I \sqrt{W_{\textrm{line}} \ \Delta X},
\label{eqn:1}
\end{eqnarray}
where $\Delta \lambda$ is the width of a resolution element (in \AA),
$\Delta I$ is the 1$\sigma$ root-mean-square fluctuation of the flux
around a normalized continuum level, $W_{\textrm{line}}$ is the full-width at
half-maximum (FWHM) of the expected spectral feature (in \AA), $B$ is
the number of bins per resolution element in the spectrum, and $\Delta
X$ is the bin size of the spectrum (in \AA).  For the spectrum shown in 
in Figure \ref{fig:halpha}, $\Delta I = $ \OneSigma{} and $\Delta X = 4$ \AA leading to $W_\lambda(3\sigma) = $ \WThreeeSigma{} for $W_{\textrm{line}} = 22$ \AA{}.

To translate this equivalent width constraint into a constraint on the
amount of material stripped from a non-degenerate companion we follow the analysis of \citet{mattila05}.  \citet{mattila05} estimate that $0.5$
\msun{} of solar-abundance material 380 days after explosion produces
an H$\alpha$ feature with peak luminosity of $\sim$3.36 $\times
10^{35}$ ergs s\ensuremath{^{-1}} \AA\ensuremath{^{-1}}.  Accounting
for the distance to M101 and
Galactic extinction towards M101 of $E(B - V) = 0.009$ mag
\citep{schlegel98} and assuming $R_{V}=3.1$, the expected H$\alpha$
peak flux from $0.5$ \msun{} of stripped material in SN 2011fe
is \PeakFluxMattila{}.  To allow for comparison with previous works, we initially approximate the H$\alpha$ feature as a Gaussian with FWHM
$22$ \AA{}.  This feature would then have an equivalent width of
$W_\lambda(0.05 \ \msun{}) = $ \WMattila{}.  Scaling linearly from the
equivalent width of the H$\alpha$ emission line to the amount of
stripped material, we place an upper limit on the amount of solar-abundance material in SN 2011fe of \MassLimit{}.

\subsection{Conservative Limit}
\label{sec:ConLimit}

Unlike the \citet{leonard07} study, the main
uncertainty in our H$\alpha$ limit arises from the continuum
determination and the H$\alpha$ line profile rather than the photon noise of our spectrum.  For example there is a small amplitude feature in the observed spectra at $6575 \-- 6595$ \AA{}.  
To test if this spectral feature could be attributed to material asymmetrically stripped from a companion we took the velocity distribution of stripped material from Figure 9 of \citet{liu12} and assumed it was narrowly distributed along a line (see Figure 11 of \citealp{liu12}).  We found that a total flux of \FluxLimitModel{} and an angle between the stripped material and our line of sight of \Angle{}\degree{} best reproduces the feature in our smoothed continuum.  To be conservative, we take a less stringent limit on the total H$\alpha$ flux of \FluxLimitPlotted{}, which is weaker than our statistical limit and too large to explain the spectral feature between $6575 \-- 6595$ \AA{} shown in Figure \ref{fig:halpha}.  This limit corresponds to an equivalent width of \WPlotted{}, a total H$\alpha$ luminosity of \FluxLimitSNPlotted{}, and \MassLimitPlotted{} of solar-abundance material linearly scaling from the models of \citet{mattila05}.  We emphasize that whether this feature is H$\alpha{}$ emission or clumpiness in the underlaying SN continuum, our conservative flux limit holds.

\section{Summary}
\label{sec:conclusion}

We obtained five deep, medium resolution, spectra of the
nearby ``plain vanilla'' \snia{} 2011fe with LBT/MODS from $73 \--
274$ days after maximum $B$-band light.  With the last nebular phase
spectrum we place the deepest flux limits on narrow H$\alpha$ emission
yet for a \snia{}. Determining the late-time H$\alpha$ emission in \sneia{} spectra requires difficult radiative transfer calculations, but linearly scaling from the models of \citet{mattila05}, our
limit translates into an upper limit on the mass of solar-abundance material of $\lesssim$\MassLimitPlotted{}, an order of a magnitude smaller than
previous limits \citep{leonard07}.  However, two important theoretical questions remain. First, the opacity of the high-velocity iron-rich ejecta should be recomputed to confirm when the SN ejecta becomes optically thin.  Second, the excitation of H$\alpha$ emission by gamma-ray deposition should be modeled in detail assuming various velocity profiles with differing amounts of stripped companion material. These issues notwithstanding, our mass limit poses a significant challenge to more exotic SD models proposed to evade previous constraints.  Additionally, our limit strongly rules out all MS and RG companions for which hydrodynamic simulations of the SN ejecta's impact are presented in the literature.

\acknowledgments

We thank Douglas C. Leonard for discussions and for providing the
nebular phase spectra of SN 2005am and SN 2005cf.  We also thank Chris
Kochanek, Todd Thompson, Jennifer van Saders, Gisella DeRosa, Dale Mudd, and Joe Antognini for
discussions and encouragement. We also thank the anonymous referee for his/her helpful suggestions. BJS was supported by a Graduate
Research Fellowship from the National Science Foundation.  
BJS and KZS
are supported in part by NSF grant AST-0908816.  
This paper used data
obtained with the MODS spectrographs built with funding from NSF grant
AST-9987045 and the NSF Telescope System Instrumentation Program
(TSIP), with additional funds from the Ohio Board of Regents and the
Ohio State University Office of Research.  
The LBT is an international collaboration among institutions in the United States, Italy and Germany. LBT Corporation partners are: The Ohio State University, and The Research Corporation, on behalf of The University of Notre Dame, University of Minnesota and University of Virginia; The University of Arizona on behalf of the Arizona university system; Istituto Nazionale di Astrofisica, Italy; LBT Beteiligungsgesellschaft, Germany, representing the Max-Planck Society, the Astrophysical Institute Potsdam, and Heidelberg University.
This research used the facilities of the Italian Center for Astronomical Archive (IA2) operated by INAF at the Astronomical Observatory of Trieste.
The SDSS is managed by the Astrophysical Research Consortium for the Participating Institutions. 
Funding for the SDSS and SDSS-II has been provided by the Alfred P. Sloan Foundation, the Participating Institutions, the National Science Foundation, the U.S. Department of Energy, the National Aeronautics and Space Administration, the Japanese Monbukagakusho, the Max Planck Society, and the Higher Education Funding Council for England. The SDSS Web Site is http://www.sdss.org/.
This research has made use of the NASA/IPAC Extragalactic Database
(NED) which is operated by the Jet Propulsion Laboratory, California
Institute of Technology, under contract with the National Aeronautics
and Space Administration.
This research has made use of NASA's Astrophysics Data System Bibliographic Services.

\bibliographystyle{apj}

\end{document}